\documentclass[a4paper,11pt]{article}
\pdfoutput=1 
\usepackage{jcappub}
\usepackage[T1]{fontenc}
\usepackage{color}
\usepackage{tcolorbox}
\usepackage{graphicx}
\usepackage{float}
\newcommand{\dd}{\text{d}}
\newcommand{\newc}{\newcommand}
\newc{\be}{\begin{equation}}
\newc{\ee}{\end{equation}}
\newc{\bal}{\begin{align}}
\newc{\eal}{\end{align}}
\newc{\ba}{\begin{eqnarray}}
\newc{\ea}{\end{eqnarray}}
\newc{\bea}{\begin{eqnarray*}}
\newc{\eea}{\end{eqnarray*}}
\newc{\D}{\partial}
\newc{\som}{\sin\omega}
\newc{\com}{\cos\omega}
\newc{\sth}{\sin\theta}
\newc{\cth}{\cos\theta}
\newc{\stom}{\sin^2\omega}
\newc{\ctom}{\cos^2\omega}
\newc{\stth}{\sin^2\theta}
\newc{\ctth}{\cos^2\theta}
\newc{\ie}{{\it i.e.} }
\newc{\eg}{{\it e.g.} }
\newc{\etc}{{\it etc.} }
\newc{\etal}{{\it et al.}}
\title{\centering
Dynamically induced\\ Planck scale and inflation\\ in the Palatini formulation 
}
\author[a]{Ioannis D. Gialamas,}
\author[b]{Alexandros Karam,} 
\author[b]{Antonio Racioppi}
\affiliation[a]{National and Kapodistrian University of Athens, Department of Physics,
Nuclear and Particle Physics Section,   GR--157 84   Athens,~Greece }
\affiliation[b]{Laboratory of High Energy and Computational Physics, National Institute of Chemical Physics and Biophysics, R\"avala 10, 10143 Tallinn, Estonia}
\emailAdd{i.gialamas@phys.uoa.gr}
\emailAdd{alexandros.karam@kbfi.ee}
\emailAdd{antonio.racioppi@kbfi.ee}

\abstract{We study non-minimal Coleman-Weinberg inflation in the Palatini formulation of gravity in the presence of an $R^2$ term. The Planck scale is dynamically generated by the vacuum expectation value of the inflaton via its non-minimal coupling to the curvature scalar $R$. We show that the addition of the $R^2$ term in Palatini gravity makes non-minimal Coleman-Weinberg inflation again compatible with observational data.} 

\begin{document}
\maketitle
\flushbottom
\section{Introduction}
In physical cosmology, cosmic inflation \cite{Starobinsky1980, Guth1981, Linde1982, Albrecht1982} is a theory which describes a period of exponential expansion of space in the early Universe. The theory of inflation manages to simultaneously solve basic issues of the Big Bang cosmology like the horizon and flatness problems. Also, it produces a power spectrum of small primordial inhomogeneities which can be compatible with the latest observational data \cite{Akrami:2018odb,Ade:2018gkx}. These data provide a very precise measurement of the scalar spectral index $n_s$ and an upper bound on the tensor-to scalar-ratio $r$\footnote{In the latest observational data~\cite{Akrami:2018odb} the spectral index has the measured value $n_s=0.9649 \pm 0.0042$ 
and the tensor-to scalar-ratio $r$ has an upper bound $r<0.056$.}. The existing data on $n_s$ and $r$ begin to impose constraints on the number of $e$-folds $N_e$, which generally depend on the reheating temperature $T_{\rm reh}$ and the equation of state parameter. In the case where rapid thermalization occurs after the end of inflation, $N_e$ does not depend anymore on $T_{\rm reh}$, which reaches its maximum value.

The use of a Coleman-Weinberg (CW) type of potential \cite{Coleman1973} to address the inflationary problem was already introduced in the early papers on inflation \cite{Linde1982, Albrecht1982, Linde1982a, Ellis1983, Ellis1983a} and has attracted the interest of many authors since \cite{Shaposhnikov2009a, GarciaBellido:2011de, Bezrukov:2012hx, Khoze2013, Kannike2014, Csaki2014, Kannike2015a, Kannike2016b, Barvinsky:2015uxa, Farzinnia2016, Rinaldi2016a, Marzola2016b, Barrie2016, Ferreira2016, Kannike2017a, Marzola2016, Karananas2016, Tambalo2017, Kannike2017, Artymowski2017, Ferreira:2016wem, Salvio2017, Karam2017, Kaneta2017, Karam2017a, Racioppi2018, Ferreira:2018a, Barnaveli:2018dxo, Kubo:2018kho, Mooij:2018hew, Shaposhnikov:2018nnm, Wetterich:2019qzx, Vicentini:2019etr, Shkerin:2019mmu, Ferreira:2019zzx, Ghilencea:2019rqj, Salvio:2019wcp, Oda:2019iwc, Racioppi:2019jsp, Benisty:2020nuu, Benisty:2020vvm, Tang:2020ovf}. In the CW models of inflation the non-minimal coupling to gravity can be added to the action (\eg\cite{Panotopoulos2014, Okada2016}) regardless of the presence of the usual $\frac{M_P^2}{2} R$ term, introducing another old concept regarding the dynamical generation of the Planck scale (\eg\cite{Cooper:1982du, Salvio2014, Racioppi2017, Azri:2018qux} and refs.~therein). In this family of models, the Planck scale is dynamically generated via the non-minimal coupling term $\frac{\xi \phi^2}{2} R$ assuming that the inflaton vacuum expectation value (VEV) $v$ is given by $v^2=\frac{M_P^2}{\xi}$. This mechanism can (but not necessarily) be related to the intriguing idea of classical scale invariance as a solution of the naturalness problem (\eg\cite{Heikinheimo2014,Gabrielli2014} and refs therein): a concept that became again quite popular few years ago, after the Higgs boson discovery. The consequences of non-minimal couplings to gravity and a dynamically induced Planck scale in inflation have been already studied in several works (\eg\cite{Kannike2015a, Kannike2016b, Racioppi2017} and refs.~therein), but their predictions for the tensor-to-scalar ratio are usually outside the range of the latest observational data. 

However, modifications of general relativity (GR) require also a discussion of what are the gravitational degrees of freedom (dof). It has been known that the Palatini formulation \cite{Sotiriou:2006hs, Sotiriou:2006qn, Sotiriou2010, Borunda:2008kf, DeFelice2010, Olmo2011, Capozziello2011, Clifton2012, Nojiri2017} of GR, is an alternative to the well-known metric formulation \cite{Padmanabhan2005, Paranjape2006}. In the Palatini formulation the connection $\Gamma_{\,\,\mu\nu}^{\lambda}$ and the metric $g_{\mu\nu}$ are treated as independent variables unlike in the metric formulation, where the connection is the Levi-Civita one. Within the context of GR the two formalisms are equivalent \cite{Exirifard2008}. However, differences between the two formulations arise when non-minimal couplings between gravity and matter are introduced \cite{Sotiriou:2006hs,Sotiriou:2006qn,Sotiriou2010} or the action no longer has a linear dependence on $R$. Perhaps the simplest modification of GR is given by adding to the Einstein-Hilbert action a function $F(R)$ \cite{Sotiriou2010} with $R$ the Ricci scalar. A notable example is the famous Starobisnky model \cite{Starobinsky1980, Mukhanov1981, Starobinskii1983} where $F(R)=\alpha R^2$. In the context of metric gravity the $R^2$ term is translated in the Einstein frame to a dynamical scalar field, the inflaton, whose mass is related to the coefficient of the $R^2$ term and gets fixed due to the observational constraint on the amplitude of the scalar perturbations $A_s \simeq 2.1 \times10^{-9}$  \cite{Akrami:2018odb,Ade:2018gkx}. Such a scenario in the metric case has attracted the interest of many authors \cite{vandeBruck:2015xpa, Kaneda:2015jma, delaCruz-Dombriz:2016bjj, Wang:2017fuy, Ema:2017rqn,Mori:2017caa, Pi:2017gih,He:2018gyf, Liu:2018hno, Gorbunov:2018llf, Elizalde:2018now, Castellanos:2018dub, Gundhi:2018wyz, Karam:2018mft, Kubo:2018kho, Enckell:2018uic, He:2018mgb, Canko:2019mud, Vicentini:2019etr}. On the other hand, any $F(R)$ theory in the framework of Palatini gravity, has no extra propagating dof that can play the role of the inflaton, therefore an additional scalar field needs to be introduced. Inflation in the context of Palatini gravity has been extensively discussed in \cite{Koivisto:2005yc,Bauer2008,Enqvist:2011qm,Borowiec:2011wd,Stachowski:2016zio,Rasanen2017,Tenkanen2017,Jarv:2017azx,Markkanen:2017tun,Enckell:2018kkc,Carrilho:2018ffi,Wu:2018idg,Kozak:2018vlp,Rasanen:2018fom,Rasanen:2018ihz,Shimada:2018lnm,Takahashi:2018brt,Jinno:2018jei,Rubio:2019ypq,Jinno:2019und,Giovannini:2019mgk,Tenkanen:2019wsd,Tenkanen:2019xzn,Bostan:2019uvv,Bostan:2019wsd,Shaposhnikov:2020geh,Shaposhnikov:2020fdv,Borowiec:2020lfx,Jarv:2020qqm,Tenkanen:2020dge,Tenkanen:2020cvw}. In particular, the impact of $F(R)=\alpha R^2$ in the Palatini gravity has received considerable attention \cite{Bombacigno:2018tyw, Enckell2019, Antoniadis2018, Antoniadis:2018yfq, Tenkanen:2019jiq, Edery:2019txq, Tenkanen:2019wsd, Gialamas:2019nly, Antoniadis:2019jnz, Antoniadis:2020dfq, Tenkanen:2020cvw, Tenkanen:2020dge, Lloyd-Stubbs:2020pvx, Ghilencea:2020piz, Das2020}. The advantage of this last class of models is that the addition of the $R^2$ term can be used to lower the tensor-to-scalar ratio $r$ \cite{Enckell2019}, in any model with a scalar field. 

In this paper we study the predictions of non-minimal CW inflation in presence of an $R^2$ term in the Palatini formulation of gravity.
The article is organized as follows: In section \ref{sec2}, we give a set up of the $R^2$ Palatini gravity in the presence of a non-minimal coupling between gravity and the inflaton. In section \ref{sec3}, we present a general discussion of the CW inflation where a dynamical generation of the Planck scale takes place due to the inflaton non-minimal coupling. In section \ref{sec4}, we present an extensive study of the inflationary phenomenology, including generalities for the end of inflation, number of $e$-folds, reheating temperature and numerical results for the models of section \ref{sec3}. We also establish an upper bound on the coefficient of the $R^2$ term $\alpha$ based on detectability of future satellites. The validity of the model is also discussed in detail. In section \ref{sec5} we end with our conclusions. In appendix \ref{appendix} we give more details about the frame transformations of section \ref{sec2}.

Throughout the paper, we use the space-time metric $\eta_{\mu\nu}=\rm{diag}(-,+,+,+)$, with Greek letters referring to space-time indices, $(0,1,2,3)$. Moreover, we have set the reduced Planck mass $M_P= (8\pi G_N)^{-1/2}$ to be dimensionless and equal to unity.

\section{The $R^2$ term in the Palatini formalism}\label{sec2}

In this section we review the main aspects of a non-minimally coupled inflaton $\phi$ in presence of an $R^2$ term in the Palatini formalism.

\subsection{The action}
Let us start by considering an action of the form~\cite{Antoniadis2018, Enckell2019}
\be \label{action1}
  S = \int\dd^4 x \sqrt{-g} \left[ \frac{\alpha}{2} R^2 + \frac{1}{2} A(\phi) R - \frac{1}{2} g^{\mu\nu} \partial_\mu \phi \partial_\nu \phi - V(\phi) \right] \,,
\ee
where $g$ is the determinant of the spacetime metric $g_{\mu\nu}$, $\alpha$ is a positive constant, $R=g^{\mu\nu} R_{\mu\nu}(\Gamma, \partial\Gamma)$ is the curvature (Ricci) scalar and $R_{\mu\nu}$ is the Ricci tensor which is built by contraction of the Riemann tensor $R^{\lambda}_{\ \, \mu\nu\sigma}$. The latter is, in turn, constructed from the connection $\Gamma$ and its first derivatives, while $\phi$ is the inflaton field and $V(\phi)$ its (Jordan frame) potential. We have also included in the action a general non-minimal function $A(\phi)$ which couples the inflaton to gravity and assumed that the inflaton kinetic term is canonical. In the following, we adopt the Palatini formulation of general relativity, where the metric $g_{\mu\nu}$ and the connection $\Gamma$ are treated as independent variables, with the extra assumption that the connection is torsion-free, $\Gamma^\lambda_{\mu\nu} = \Gamma^\lambda_{\nu\mu}$.

In order to obtain a minimally coupled inflaton field, we eliminate the $R^2$ term by introducing a non-minimally coupled auxiliary field $\chi \equiv  2 \alpha R$ and we perform a Weyl transformation (which depends on both $\phi$ and $\chi$) \cite{Enckell2019}
\be
  g_{\mu\nu} \rightarrow \Omega^2 g_{\mu\nu} = [ \chi + A(\phi) ] g_{\mu\nu} \ . \label{eq:Weylscaling}
\ee
Then, the action becomes
\be \label{action4}
  S = \int\dd^4 x \sqrt{-g} \left[ \frac{1}{2} R - \frac{1}{2} \frac{1}{\chi + A(\phi)} g^{\mu\nu} \partial_\mu \phi \partial_\nu \phi - \hat V(\phi, \chi) \right] \ ,
\ee
where the conformally transformed potential is
\be \label{U2}
  \hat V(\phi, \chi) = \frac{1}{[\chi + A(\phi)]^2} \left[ V(\phi) + \frac{\chi^2}{8 \alpha} \right] \ .
\ee
Varying \eqref{action4} with respect to $\chi$, we obtain a constraint equation with the solution
\be \label{chi}
  \chi = \frac{ 8 \alpha V(\phi) + 2 \alpha A(\phi) \left( \partial \phi \right)^2 }{ A(\phi) - 2 \alpha \left( \partial \phi \right)^2 } \ .
\ee
In general, we can insert \eqref{chi} into \eqref{U2} to eliminate $\chi$ and write the action in terms of $\phi$ only. We obtain
\be \label{action5}
  S = \int\dd^4 x \sqrt{-g} \left[ \frac{1}{2} R + \frac{1}{2} 	K(\phi) \left( \partial \phi \right)^2 + \frac{1}{4}L(\phi)  \left( \partial \phi \right)^4  - \frac{\bar{U}}{1+8\alpha \bar{U}}  \right] \ ,
\ee
where we have defined
\be \label{Ub}
  \bar{U}(\phi) \equiv \frac{V(\phi)}{[A(\phi)]^2} 
\ee
and
\be \label{KandL}
  K(\phi) \equiv -\frac{1}{ A (1+8\alpha \bar{U}) }\,, \qquad L(\phi)=\frac{2\alpha}{A^2 (1+8\alpha \bar{U})}\,. 
\ee
Note that $\bar{U}$ is the usual Einstein frame potential in the case where we do not have the $R^2$ term (for more details see Appendix \ref{appendix}). Also, in addition to modifying the potential, the conformal transformation has translated the $R^2$ term into a higher-order kinetic term for the inflaton field, which is always subdominant during slow-roll \cite{Enckell2019}. Finally, through a field redefinition of the form
\be 
  \left(\frac{\dd \phi}{\dd \zeta}\right)^2 = A ( 1 + 8 \alpha \bar{U} ) \ , \label{eq:phi:general}
\ee
the action takes the form
\be \label{action6}
  S = \int\dd^4 x \sqrt{-g} \left[ \frac{1}{2} R - \frac{1}{2} \left( \partial \zeta \right)^2 + \frac{\alpha}{2} \left( 1 + 8 \alpha \bar{U}(\zeta) \right) \left( \partial \zeta \right)^4 - U(\zeta)  \right] \,,
\ee
where the Einstein frame potential reads
\be \label{Ueff}
  U \equiv \frac{\bar{U}}{1 + 8 \alpha \bar{U}}\,. 
\ee
We can see that, regardless of the shape of $\bar{U}$, the $R^2$ term decreases the height of the Einstein frame potential. For large field values, the effective potential becomes flat and asymptotes to the value $1/(8\alpha)$ \cite{Enckell2019}.

\section{Coleman-Weinberg potential}\label{sec3}
In this section we discuss the details about the inflaton potential, first for the original models with $\alpha=0$ and later also in the presence of $\alpha$. We consider the following scalar potential
 \begin{equation}
  V(\phi) = \frac{1}{4} \lambda (\phi) \phi^4 + \Lambda^4 \label{eq:Veff:J},
 \end{equation}
 containing a running\footnote{The careful reader might notice that also $\xi$ and $\alpha$ are subject to quantum corrections. However, it can be proven that their running is suppressed and can be safely ignored because of the constraint on the amplitude of scalar perturbations \cite{Akrami:2018odb, Ade:2018gkx} and perturbativity of the theory (\eg\cite{Salvio2014, Marzola2016, Markkanen:2018bfx, Vicentini:2019etr} and refs.~therein.)} quartic coupling $\lambda (\phi)$ and a cosmological constant $\Lambda$ which is adjusted so that at the minimum the potential value is zero, \ie,
 \begin{equation}
  V(v)=\frac{1}{4} \lambda (v) v^4 + \Lambda^4 = 0 \label{eq:Vmin} \, ,
 \end{equation}
 where $v$ is the VEV of the inflaton.  We assume the following non-minimal coupling to gravity:
\begin{equation}
  A(\phi) = \xi \phi^2 \,,
  \label{eq:A}
\end{equation} 
therefore the action \eqref{action1} lacks an Einstein-Hilbert term, which is dynamically generated by a non vanishing inflaton VEV that satisfies
\begin{equation}
  v = \frac{1}{\sqrt\xi}.
  \label{eq:v:phi:Planck:mass}
\end{equation}
Note that such a relation automatically implies that $\xi$ can only take positive values. 
We discuss now the possible scenarios that arise from the minimization of the scalar potential. A complete discussion was already presented in \cite{Racioppi2017}, however for the sake of clarity we review the relevant details. Given the scalar potential in eq.~\eqref{eq:Veff:J}, the general minimum equation is
\begin{equation}
 \frac{1}{4} \beta (v) + \lambda (v) =0 \, ,
\end{equation}
where $\beta (\mu)= \mu \frac{\partial}{\partial \mu}\lambda (\mu)$ is the beta-function of the quartic coupling $\lambda (\mu)$.
Therefore, several possibilities are open according to how we solve the equation:
\begin{eqnarray}
\text{a)} & \beta (v)=\lambda (v)=0 , \label{eq:RGE:MPCP}\quad\\
\text{b)} & \beta (v) > 0, \ \lambda (v)<0 ,\label{eq:RGE:bound:cond}\\
\text{c)} & \beta (v) < 0, \ \lambda (v)>0  .
\end{eqnarray}
It is easy to show that c) is actually a local maximum of the potential, therefore the only allowed solutions are a) or b). Using eq.~\eqref{eq:Vmin}, the first option implies also that $\Lambda=0$, realizing a full classical scale invariant setup, while the second option requires $\Lambda \neq 0$ (it can be proven that scale invariance is only softly broken \ie $\Lambda \ll 1$ \cite{Racioppi2017}).
The quartic coupling pre-factor in eq.~\eqref{eq:Veff:J} can be model-independently written as a Taylor expansion around the VEV
\begin{equation}
\lambda(\phi) = \lambda(v) + 
                                  \beta(v) \ln\frac{\phi }{v}+
                  \frac{1}{2!} \beta'(v) \ln^2\frac{\phi }{v} +
                     \frac{1}{3!} \beta ''(v)  \ln^3\frac{\phi }{v} + \cdots ,
  \label{eq:lambdaTaylor}
\end{equation}
where $\beta'(\mu)$ and $\beta ''(\mu)$ are respectively the first and second derivative of $\beta(\mu)$ with respect to $t=\ln\mu$ and we assumed without loss of generality that $\phi>0$. Therefore for case a) described in eq.~\eqref{eq:RGE:MPCP} we have that the leading order expression is 
\begin{equation}
\lambda^a(\phi) \simeq  \frac{ \beta'(v)}{2} \ln^2\frac{\phi }{v} \, , \label{eq:lambda:run:2}
\end{equation}
 while for case b) we get
\begin{equation}
\lambda^b(\phi) \simeq \lambda(v) + \beta(v) \ln\frac{\phi    }{v} \, . \label{eq:lambda:run}
\end{equation}
In the following subsections we discuss separately each case, starting from case b). In order to avoid a cumbersome notation, from now on we omit the argument ``$(v)$'' and restore it only when needed.

\subsection{1st order Coleman-Weinberg potential} \label{subsec:xi:CW}
By using eqs.~\eqref{eq:Vmin}, \eqref{eq:v:phi:Planck:mass} and \eqref{eq:lambda:run} the potential can be rewritten as \cite{Kannike2016b, Racioppi2017}
\begin{eqnarray}
  V(\phi)&=& \Lambda ^4 \left\{ 1 + \left[ 4 \ln \left(\frac{\phi}{v}\right) -1 \right] \frac{\phi^4}{v^4} \right\} \label{eq:V:CW}
\label{eq:Veff:Jordan:Lambda}.
\end{eqnarray}
In presence of the non-minimal coupling to gravity (\ref{eq:A}) but in absence of an $R^2$ term, the inflaton potential in the Einstein frame becomes \cite{Kannike2016b, Racioppi2017}
\begin{equation}
{\bar U}({\bar \zeta})=\Lambda ^4 \left(4 \, \frac{{\bar \zeta}}{v}+e^{-4 \,  \frac{{\bar \zeta}}{v}}-1\right) \, , \label{eq:V:CW:Einstein:0}
\end{equation}
where the field redefinition is
\begin{equation}
 \phi = e^{\bar \zeta/v} v \, .    \label{eq:phi} 
\end{equation}
We can immediately appreciate two relevant limit cases \cite{Kannike2016b,Racioppi2017}. For  $v \ll 1$ (\ie $\xi \gg 1$) and ${\bar \zeta} > 0$, the potential becomes
\begin{equation}
{\bar U}({\bar \zeta}) \approx a_\zeta \, {\bar \zeta}  \, , \label{eq:V:CW:linear}
\end{equation}
with $a_\zeta = 4 \frac{\Lambda ^4}{v}$. 
On the other hand for $v \gg 1$ (\ie $\xi \ll 1$), the potential reduces to
\begin{equation}
{\bar U}({\bar \zeta}) \approx \frac{m^2}{2} {\bar \zeta}^2 \, , \label{eq:V:CW:quadratic}
\end{equation}
with $m = m_1 = 4 \frac{\Lambda^2}{v}$. Therefore in absence of $\alpha$, the model includes linear and quadratic inflation as limit solutions respectively for small $v$ and big $v$. Let us see now how the predictions change with the addition of the $R^2$ term. In this case the Einstein frame potential becomes\footnote{More details on the use of the $\bar \zeta$ field redefinition in presence of an $R^2$ term are given in the Appendix.} 
\begin{equation}
 U(\bar \zeta) =  \frac{ \Lambda ^4 \left(4 \, \frac{{\bar \zeta}}{v}+e^{-4 \,  \frac{{\bar \zeta}}{v}}-1\right)  }{1 + 8 \alpha \Lambda^4 \left(4 \, \frac{{\bar \zeta}}{v}+e^{-4 \,  \frac{{\bar \zeta}}{v}}-1\right)} \, ,
\end{equation}
where we have used eq.~\eqref{Ueff} with $\bar U$ given in  \eqref{eq:V:CW:Einstein:0} and $A(\bar \zeta) = 1$. Unfortunately in this case the corresponding field redefinition \eqref{eq:phi:general}, $d \zeta/d \bar \zeta$, cannot be solved exactly. However, we can still present exact solutions for the two limit cases shown above. In such cases the result for $U$ is equivalent to the solution of the model with a minimally coupled inflaton with a linear/quadratic scalar potential plus an $R^2$ term. We obtain that
\begin{equation}
    U({\zeta}) \simeq \frac{1}{8 \alpha} 
    \frac{1+\frac{1}{2 \alpha  a_\zeta \zeta}}{\left(1+\frac{1}{4 \alpha  a_\zeta \zeta}\right)^2} \,
    \label{eq:U:CW:small:v}
\end{equation}
for small $v$, where the field the redefinition is very well approximated with
\begin{equation}
 {\bar \zeta} \simeq \zeta + 2 \alpha  a_\zeta \zeta ^2    \, .
\end{equation}
On the other hand, for  big $v$, we get
\begin{equation}
   U({\zeta}) \simeq  \frac{ \tanh ^2\left(2 \sqrt{\alpha } m  \zeta\right)}{8 \alpha } \label{eq:U:tanh} \, 
\end{equation}
where now the field the redefinition is very well approximated by
\begin{equation}
 {\bar \zeta} \simeq \frac{1 }{2 \sqrt{\alpha } m} \sinh \left(2 \sqrt{\alpha } m \zeta\right)  \, .
\end{equation}
This last result is in agreement with \cite{Tenkanen:2019wsd}. However, with respect to \cite{Tenkanen:2019wsd}, our study will present an improved study of the reheating phenomenology and the computation of the number of $e$-folds (see section \ref{sec4}). In order to have an understanding of the overall shape of the potential, in figure \ref{fig:U:vs:zetabar}, red line, we plot the 1st order CW potential as a function of $\bar \zeta$ for the reference values $\xi = 10$, $\Lambda = 0.0015$. In the left panel we show $\bar{U}$, \ie the potential for $\alpha = 0$, while on the right panel we show $U$, \ie the  potential in the presence of the $\alpha R^2$ term, with  $\alpha = 10^{10}$. We can notice that since ${\bar U}({\bar \zeta})$ is asymmetrical under the transformation ${\bar \zeta} \to -{\bar \zeta}$, the same holds for $U$. However, in both the quadrants the asymptotic limit is $U \to \frac{1}{8 \alpha}$ in agreement with \cite{Enckell2019}.

\subsection{2nd order Coleman-Weinberg potential} \label{subsec:xi:CW:2}
\begin{figure}[t]
\begin{center}
\includegraphics[width=0.495\textwidth]{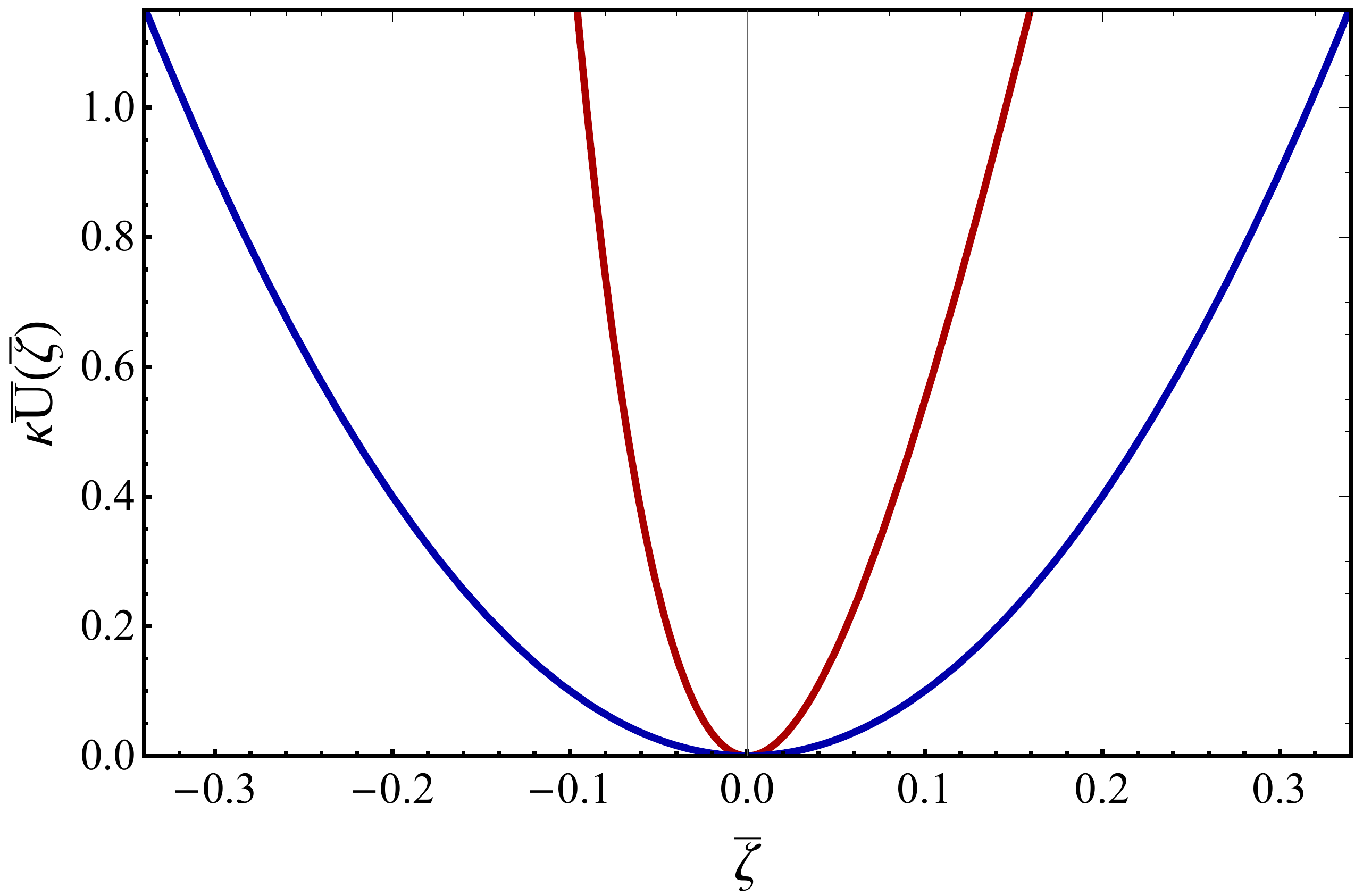}
\includegraphics[width=0.495\textwidth]{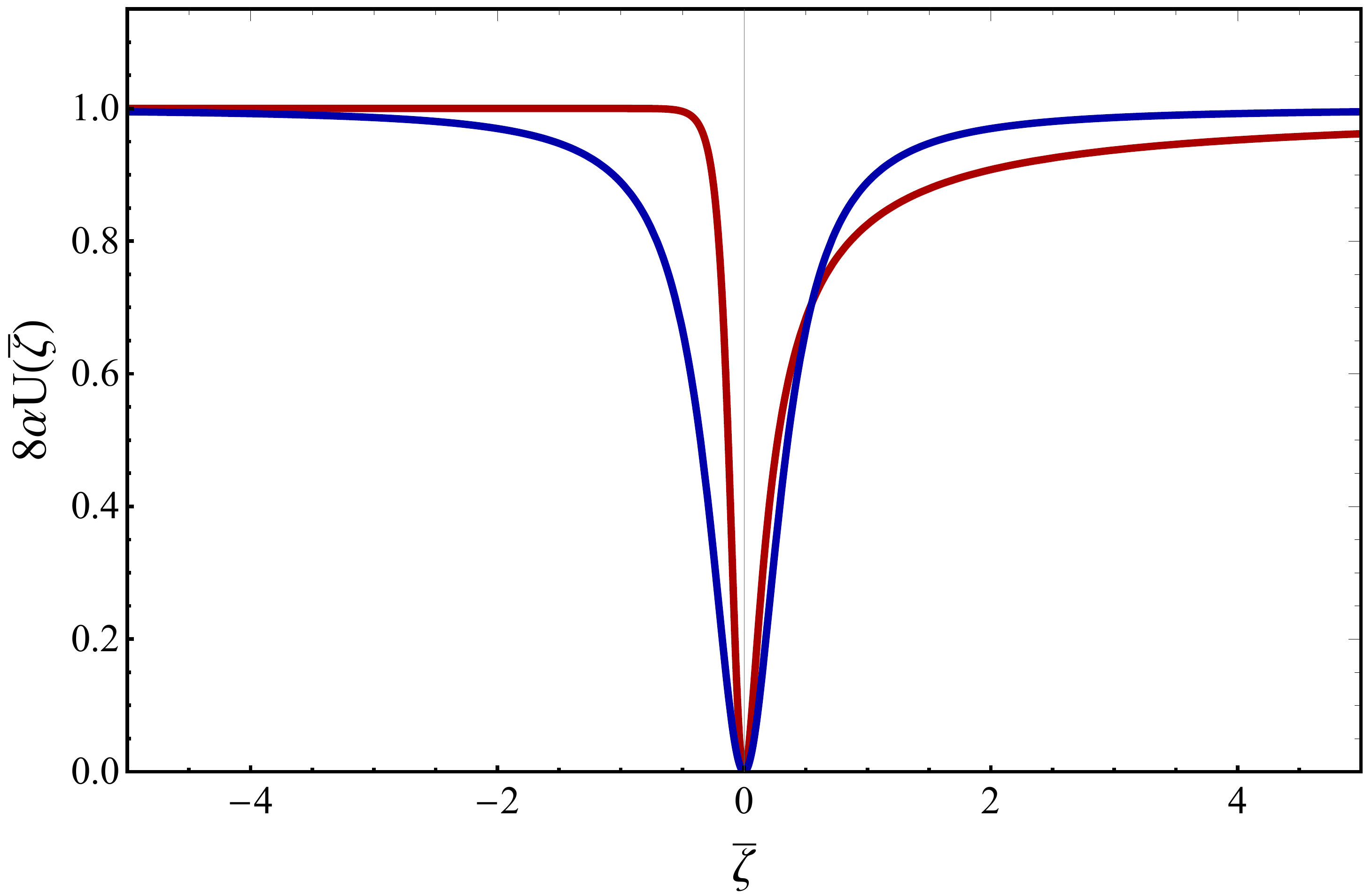}
\caption{\sf \textbf{Left:} Scalar potential $\bar{U}(\bar{\zeta})$ for the 1st order CW model in section \ref{subsec:xi:CW} (red line) with $\kappa = \Lambda^{-4}$ and for the 2nd order CW model in section \ref{subsec:xi:CW:2} (blue line) with $\kappa = \xi^2/ \beta'$. For both potentials we fixed $\xi = 10$. \textbf{Right:} Scalar potential $U(\bar{\zeta})$ for the 1st order CW model in section \ref{subsec:xi:CW} for $\Lambda = 0.0015$ (red line) and for the 2nd order CW model in section \ref{subsec:xi:CW:2} for $ \beta' = 10^{-9}$ (blue line). For both potentials we fixed $\xi = 10$ and $\alpha = 10^{10}$. 
}
\label{fig:U:vs:zetabar}
\end{center}
\end{figure}

By using eqs.  (\ref{eq:Vmin}), (\ref{eq:v:phi:Planck:mass}) and (\ref{eq:lambda:run:2}) the potential can be rewritten as \cite{Kannike2015a}
\begin{equation}\label{potCW2}
    V(\phi) = \frac{1}{8}  \beta'  \phi ^4 \ln ^2\left(\frac{\phi }{v}\right)
\end{equation}
and the non-minimal coupling satisfies again eq.~\eqref{eq:v:phi:Planck:mass}.
Without the $R^2$ term, the model reproduces in the Einstein frame the quadratic inflaton potential~\eqref{eq:V:CW:quadratic}, where now \cite{Kannike2015a}
\begin{equation}
m^2 = m^2_2 = \frac{ \beta'  v^2}{4} \, . \label{eq:m:CW:2}
\end{equation}
Therefore the inflaton potential in the presence of the $R^2$ term is again \eqref{eq:U:tanh} with the mass parameter provided by eq.~\eqref{eq:m:CW:2}. 
In figure \ref{fig:U:vs:zetabar}, blue line, we plot the 2nd order CW potential as a function of $\bar \zeta$ for the reference values $\xi = 10$ and $ \beta' = 10^{-9}$. In the left panel we show $\bar{U}$, \ie the potential for $\alpha = 0$, while on the right panel we show $U$, \ie the  potential in the presence of the $\alpha R^2$ term, with  $\alpha = 10^{10}$. We can notice that since ${\bar U}({\bar \zeta})$ is now symmetrical under the transformation ${\bar \zeta} \to -{\bar \zeta}$, the same holds for $U$. We can also appreciate that the asymptotic limit of $U \to \frac{1}{8 \alpha}$ holds for both models regardless of the starting potential $\bar U$ (provided that their asymptotic limit is $\bar U \rightarrow \infty$), in agreement with \cite{Enckell2019}.

\section{Inflationary phenomenology}\label{sec4}

\subsection{General equations about slow-roll}

In the flat FRW case with zero spatial curvature the Friedmann and Klein-Gordon equations for the redefined inflaton field read~\cite{Enckell2019}
\ba 
  3 H^2 &=& \frac{1}{2} [ 1 + 3 \alpha ( 1+8\alpha \bar{U} ) \dot\zeta^2 ] \dot\zeta^2 + U 
  \label{FRW1}
  \\  
  0 &=& [ 1+ 6 \alpha ( 1 + 8 \alpha \bar{U} ) \dot\zeta^2 ] \ddot\zeta + 3 [ 1 + 2 \alpha ( 1 + 8 \alpha \bar{U} ) \dot\zeta^2 ] H \dot\zeta + 12 \alpha^2 \dot\zeta^4 \bar{U}' + U' \,.
  \label{FRW2}
\ea
Inflation takes place when the first Hubble slow-roll parameter
\be 
\epsilon_H \equiv -\frac{\dot{H}}{H^2} = \frac{\dot{\zeta}^2}{2 H^2} \left[ 1 + 2 \alpha \left( 1 + 8 \alpha \bar{U} \right) \dot{\zeta}^2 \right]
\ee
is smaller than one. In the slow-roll limit, $\epsilon_H \ll 1$ and the $\ddot\zeta$ term in~\eqref{FRW2} is negligible. Analogously, the contribution of the higher-order kinetic term in~\eqref{action6} is subdominant and can be neglected (see footnote 1 of~\cite{Enckell2019}). The dynamics of the higher-order kinetic terms are usually encoded in the speed of sound parameter, which has been shown in similar Palatini-$R^2$ models to be very close to unity during inflation and only deviate at the end of it, without significantly affecting the inflationary predictions though~\cite{Gialamas:2019nly}. Also, even for large values of $\alpha$, the slow-roll attractor is reached exponentially fast~\cite{Tenkanen:2020cvw}, which means there is no need for fine tuning. Furthermore, the first order expressions for the amplitude of the scalar power spectrum and spectral index as functions of $\phi$ do not depend explicitly on $\alpha$  \cite{Enckell2019}
\be \label{eq:As-and-ns}
24 \pi^2 A_s = \frac{U}{\epsilon_U} = \frac{\bar{U}}{\epsilon_{\bar{U}}} \,, \qquad n_s = 1 - 6 \epsilon_U + 2 \eta_U = 1 - 6 \epsilon_{\bar{U}} + 2 \eta_{\bar{U}}\,,
\ee
with the potential slow-roll parameters defined as
\be 
\epsilon_U = \frac{1}{2} \left( \frac{U'}{U} \right)^2 \,, \qquad \epsilon_{\bar{U}} = \frac{1}{2} \left( \frac{\bar{U}'}{\bar{U}} \right)^2 = \epsilon_U \vert_{\alpha = 0}\,,
\ee
\be 
\eta_U =  \frac{U''}{U} \,, \qquad \eta_{\bar{U}} =  \frac{\bar{U}''}{\bar{U}}  = \eta_U \vert_{\alpha = 0}\,.
\ee
On the other hand, the tensor power spectrum depends explicitly on $\alpha$
\be 
A_T = \frac{2}{3 \pi^2} U = \frac{2}{3\pi^2} \frac{\bar{U}}{1 + 8 \alpha \bar{U}}\,.
\ee
Consequently, the tensor-to-scalar ratio becomes 
\be 
r = 16 \epsilon_U = \frac{\bar{r}}{1 + 8 \alpha \bar{U}}  = \frac{\bar{r}}{ 1+ 12 \pi^2 A_s \bar{r} \alpha } \,, \label{eq:r}
\ee
where in the last equality we used eq. \eqref{eq:As-and-ns} and $\bar{r}=16\epsilon_{\bar{U}}$ is the tensor-to-scalar ratio of the same model but without the $R^2$ term. 
We can see then that for large enough $\alpha$ we can lower the value of $r$ in a given model, without affecting the prediction for $n_s$ (assuming that the number of $e$-folds is independent on $\alpha$, which will turn out to be a rough approximation). 

At this point, we would like to establish an upper bound on $\alpha$ based on detectability from future experiments. The next-generation CMB satellites (LiteBIRD \cite{Matsumura2016}, PIXIE \cite{Kogut_2011}, PICO \cite{Hanany:2019lle}), if approved, will be able to detect the primordial CMB B-mode polarization for tensor-to-scalar ratio values of $r<0.001$. In particular the sensitivity of PICO will be approximately $\delta_r \approx 10^{-4}$. We use this last value to set an upper bound $\alpha$. 
For $\alpha \to \infty$ (and $\bar r > \delta_r$) we obtain from \eqref{eq:r}
\begin{equation}
     r_\text{limit} \approx \frac{1}{ 12 \pi^2 A_s \alpha } \label{eq:rlimit}
\end{equation}
Therefore setting the upper bound  $\alpha < 4 \times 10^{10}$, we ensure the possibility to test our models' predictions \ie $ r_\text{limit} > \delta_r$.

\subsection{Analytical results}
Assuming standard slow-roll, the equations for the inflationary parameters can be easily derived applying eqs.~\eqref{eq:As-and-ns} and \eqref{eq:r} to the models described in sections \ref{subsec:xi:CW} and \ref{subsec:xi:CW:2}. For what concerns the first model we cannot provide analytical solutions that are valid for all the range of parameters. However, we can provide the solutions for the two limit cases related to linear and quadratic inflation. For what concerns the linear inflation limit, the results are
        \ba 
            r &=& \frac{4}{N_e + 8 \sqrt{2}\, \alpha\, a_\zeta N_e^{3/2}} \,, \\
            n_s &=& 1 - \frac{3}{2 N_e} \,, \\
            A_s &=& \frac{a_\zeta N_e^{3/2}}{3 \sqrt{2}\, \pi^2} \,,
        \ea
while for the quadratic inflation limit we have
        \ba 
            r &=& \frac{8}{N_e + 16 \, \alpha \, m^2 N_e^2} \,, \\
            n_s &=& 1 - \frac{2}{N_e} \,, \\
            A_s &=& \frac{m^2 N_e^2}{6 \pi^2} \,,
        \ea
with $m$ referring to either $m_1$ (1st order CW potential, eq.~\eqref{eq:V:CW:quadratic}) or $m_2$ (2nd order CW potential, eq.~\eqref{eq:m:CW:2}).
However, the previous equations 
do not take into account the reheating process, which would help us to better estimate $N_e$ and introduce a dependence on $\alpha$ for $N_e$. This is investigated in the remaining part of the article.

\subsection{End of inflation}
Unfortunately, it is not always possible to analytically solve the field redefinition \eqref{eq:phi:general}. In such cases it is more convenient to work with the action \eqref{action5} and use $\phi(t)$ as dynamical variable.
In a flat FRW metric where $ \phi=\phi(t) $, the energy density and pressure are given by
\be \label{dens}
  \rho(\phi)=K(\phi)X + 3L(\phi) X^2 + U(\phi)\,,
\ee
\be \label{press}
  p(\phi)=K(\phi)X + L(\phi) X^2 - U(\phi)\,,
\ee
where  $ X=\frac{1}{2} (\partial \phi)^2 =-\frac{1}{2}\dot{\phi}^2$ and $K$, $L$ and $U$ are respectively given in eqs. \eqref{KandL} and \eqref{Ueff}.
The end of inflation is defined from the equation $ \epsilon_H = - \frac{\dot{H}}{H^2} =1\,, $ or equivalently $ \rho=-3p\,, $ which drives us to the quadratic equation \cite{Gialamas:2019nly}
\be \label{endinfeq}
  3 L(\phi_{\rm end}) X_{\rm end}^2 + 2 K(\phi_{\rm end}) X_{\rm end} -U(\phi_{\rm end})=0\,.
\ee
Because X is negative by definition, the only acceptable solution of \eqref{endinfeq} is
\be \label{Xend}
X_{\rm end}=\frac{-K(\phi_{\rm end})-\sqrt{K(\phi_{\rm end})^2+3L(\phi_{\rm end})U(\phi_{\rm end})}}{3L(\phi_{\rm end})}\,.
\ee
The full kinetic term can be written in the form 
\be \label{fullKin}
K(\phi) X\,(1+\delta_X)\,,
\ee
with 
\be \label{deltaX}
\delta_X =\frac{L(\phi)}{K(\phi)}X\,.
\ee
Substituting \eqref{Xend} in \eqref{deltaX} we obtain that at the end of inflation
\be \label{deltaXend}
\delta_{X_{\rm end}} =\frac{-1+\sqrt{1+3L(\phi_{\rm end})\,U(\phi_{\rm end})/K(\phi_{\rm end})^2}}{3}\,.
\ee
It is a fairly easy task to see from \eqref{deltaXend} that in the absence of the higher order in the velocity terms, $ \delta_X $ vanishes as it should be. In this way we have calculated the velocity squared of the field as a function of the field value $\phi_{end}$ at the end of inflation, which is given by the condition $\epsilon_H =1$. At the first order in the slow-roll parameters, this is achieved when \cite{Ellis2015a}
\be \label{endinf}
 \epsilon_U \simeq \left(1 + \sqrt{1 - \frac{\eta_U}{2}} \right)^2 \,,
\ee
which we solve numerically in order to determine $\phi_{\rm end}$.

\subsection{Reheating temperature and number of $e$-folds}

In this section we do not propose a particular model of reheating, but we will calculate the instantaneous reheating temperature of the 1st and 2nd order CW models studied  in the previous sections. The reheating temperature in various models and mechanisms has been extensively discussed~\cite{Kofman1994,Kofman1997,Liddle:2003as,Dodelson2003,Podolsky:2005bw,Martin:2010kz,Allahverdi:2010xz,Adshead:2010mc,Mielczarek:2010ag,Easther:2011yq,Dai2014,Munoz2015,Cook:2015vqa,Gong:2015qha,Rehagen:2015zma,Lozanov:2016hid,Lozanov:2017hjm}\footnote{See also \cite{Bassett2006} for a review on reheating mechanisms,~\cite{Kawasaki:1999na,Choi:2017ncz,Hasegawa:2019jsa} for lower bounds on the reheating temperature, \cite{Giovannini:2019oii} for reheating and relic gravitons and \cite{German:2020dih} for a recent work on constraints on the reheating temperature in inflationary models.}. Minimal models of reheating in the non-minimal CW model, both in metric and Palatini formulation of gravity have been studied in \cite{Kannike2016b} and \cite{Racioppi2017}. The reheating in Palatini $ R^2 $ models has been studied in \cite{Tenkanen:2019jiq, Gialamas:2019nly, Lloyd-Stubbs:2020pvx}.
The reheating temperature is given by
\be
T_{\rm reh}=\left(\frac{30}{\pi^2}\frac{\rho_{\rm reh}}{g^{*}_{\rm reh}}\right)^{1/4}\,,
\ee
where $ \rho_{\rm reh} $ is the energy density when  the universe becomes thermalized and $ g^{*}_{\rm reh} $ are the effective energy dof at the temperature $ T_{\rm reh}$. Assuming the standard model content, $ g^{*}_{\rm reh} $ varies from $3.36$ at $ T_{\rm reh} \sim 10\, \rm keV $ to $106.75$ at $ T_{\rm reh} \sim 1\, \rm TeV $ or higher~\cite{Husdal:2016haj}. For our studies we will adopt the highest value for the effective energy dof  $ g^{*}_{\rm reh}=106.75 $. Such an assumption will be justified by the numerical results that we will obtain (see figure \ref{fig:Trehxi}).

We suppose a rapid reheating period, where the number of $e$-folds during that is equal to zero and therefore the energy density at the end of inflation $ \rho_{\rm end}\,, $ is equal to the reheating energy density $\rho_{\rm reh} $. This temperature is called instantaneous reheating temperature and is given by
\be\label{treh}
T_{\rm reh}^{\rm ins}=\left(\frac{30}{\pi^2}\frac{\rho_{\rm end}}{g^{*}_{\rm reh}}\right)^{1/4}\,. 
\ee
As already mentioned in the previous sections, the end of inflation is  determined by  $ \epsilon_H = 1 $, thus substituting \eqref{Xend} in \eqref{dens} we can compute the energy density at the end of inflation and therefore the instantaneous reheating temperature, $ T_{\rm reh}$.\footnote{From now on, in order to speed up notation we have denoted $T_{\rm reh}^{\rm ins}=T_{\rm reh} $.} 
Moreover, in the hypothesis of instantaneous reheating, the number of $e$-folds can be very well approximated as \cite{Ellis2015a} 
\begin{equation}
N_e \simeq 61.1 + 
   \frac{1}{4} \ln
   \left(\frac{U_*^2}{\rho_{\rm end} }\right) \, , \label{eq:Ne}
\end{equation}
where $\rho_{\rm end}$ is the energy density at the end of inflation, which can be computed using eqs.~\eqref{dens} and \eqref{Xend}, as discussed before.
Performing the computations we obtain 
\begin{equation}
N_e \simeq 61.1 + \frac{1}{4} \ln \left(\frac{6 \alpha  \bar{U}_*^2 \left(1+8 \alpha  \bar{U}_{\rm end}\right)}{
   \left(1+8 \alpha  \bar{U}_*\right)^2 \left(1-\sqrt{1+6 \alpha  \bar{U}_{\rm end}}+12 \alpha 
   \bar{U}_{\rm end}\right)}\right) \label{eq:Ne:2}
\end{equation}
and taking the leading order term for $\alpha \to \infty$ we get
\begin{equation}
 N_\text{limit} \simeq 60.4 - \frac{1}{4} \ln \alpha \label{eq:Ne:3}
\end{equation}
Therefore the number of $e$-folds is decreasing with $\alpha$ increasing in agreement with \cite{Lloyd-Stubbs:2020pvx}. However, the numerical part of our results is slightly different because in \cite{Lloyd-Stubbs:2020pvx} an extra $\ln(2\pi)$ factor has been added and the assumption that the energy density during inflation is equal to $\rho_{\rm end}$  has been used. 
 Applying the detectability upper limit on $\alpha$ from \eqref{eq:rlimit} we obtain the lower limit on the number of $e$-folds to be $N _e \gtrsim 54.3$.

\subsection{Numerical results}
Next, let us compute the inflationary predictions and the reheating temperature for the 1st and 2nd order CW potentials. In all results we use $v$ as a free parameter and choose $\xi$ so that  \eqref{eq:v:phi:Planck:mass} is satisfied. Moreover, we point out that the amplitude of the power spectrum constraint $A_s \simeq 2.1 \times 10^{-9}$  \cite{Akrami:2018odb,Ade:2018gkx}, has been taken into account in all the presented results.

\subsubsection{Inflationary predictions}
\begin{figure}[t]
\begin{center}
\includegraphics[width=0.49\textwidth]{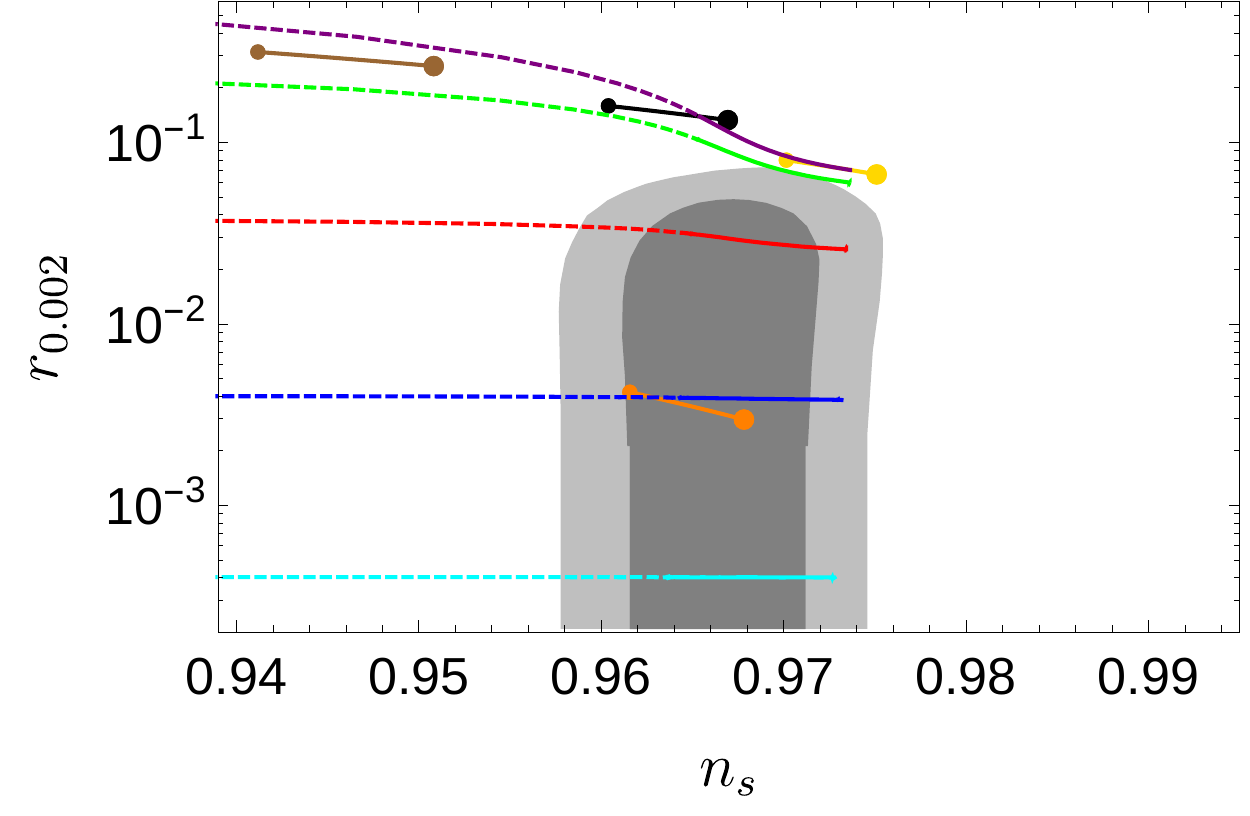}
\includegraphics[width=0.49\textwidth]{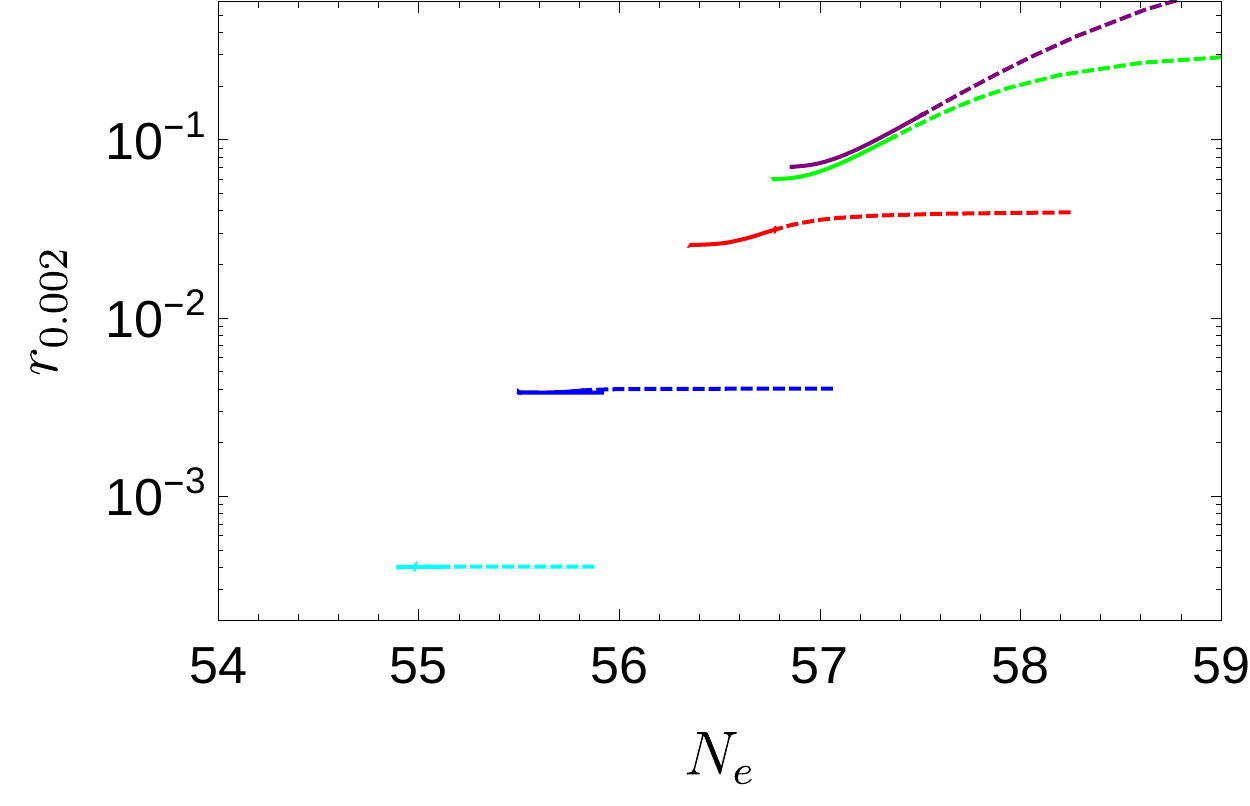}
\caption{\sf $r$ vs. $n_s$ (left) and $r$ vs. $N_e$ for the 1st order CW potential with $R^2$ in Palatini gravity with $\alpha=0$ (purple), $10^7$ (green), $10^8$ (red), $10^9$ (blue) and $10^{10}$ (cyan). The continuous line represents positive field ($\bar\zeta >0$ \ie $\phi > v$) inflation, while the dashed one negative field ($\bar\zeta <0$ \ie $\phi < v$) inflation. For reference, we also plot the predictions of quartic (brown), quadratic (black), linear (yellow) and Starobinsky (orange) inflation in metric gravity.  The light gray areas present the 1 and 2$\sigma$ constraints from the latest Planck data.
}
\label{fig:r:vs:ns:xi:CW:R2}
\end{center}
\end{figure}

For the 1st order CW potential the corresponding predictions are given in figure \ref{fig:r:vs:ns:xi:CW:R2}. In the left panel of figure \ref{fig:r:vs:ns:xi:CW:R2} we present the results for the tensor-to-scalar ratio $r$ vs. the spectral index $n_s\,.$  The continuous lines represent positive field ($\bar\zeta >0$ \ie $\phi > v$) inflation, while the dashed negative field ($\bar\zeta <0$ \ie $\phi < v$) inflation. The different colors refer to the various values of the parameter $\alpha$, namely $\alpha=0$ (purple), $10^7$ (green), $10^8$ (red), $10^9$ (blue) and $10^{10}$ (cyan) (in agreement with the \emph{assumed} upper limit due to eq.~\eqref{eq:rlimit}). The light gray areas present the 1 and 2$\sigma$ constraints from the latest Planck data  \cite{Akrami:2018odb}. For reference, we also plot the predictions of quartic (brown), quadratic (black), linear (yellow) and Starobinsky (orange) inflation in metric gravity for $N_e \in [50,60]$. As can be seen, the purple line passes through the predictions for linear and quadratic inflation in agreement with the asymptotic limits \eqref{eq:V:CW:linear} and \eqref{eq:V:CW:quadratic} \cite{Racioppi2017}. Due to \eqref{eq:r}, as the parameter $\alpha$ increases the predictions for $r$ come back into agreement with the observational data. In the right panel of figure \ref{fig:r:vs:ns:xi:CW:R2} we present the results for the tensor-to-scalar ratio $r$ vs. the number of $e$-folds $N_e\,,$ computed from \eqref{eq:Ne:2}. The line color and style is the same as before. It is worth noting that as $\alpha$ gets larger $N_e$ decreases in agreement with the leading order result \eqref{eq:Ne:3}. Consequently a similar behavior holds for $n_s$.

\begin{figure}[t]
\begin{center}
\includegraphics[width=0.49\textwidth]{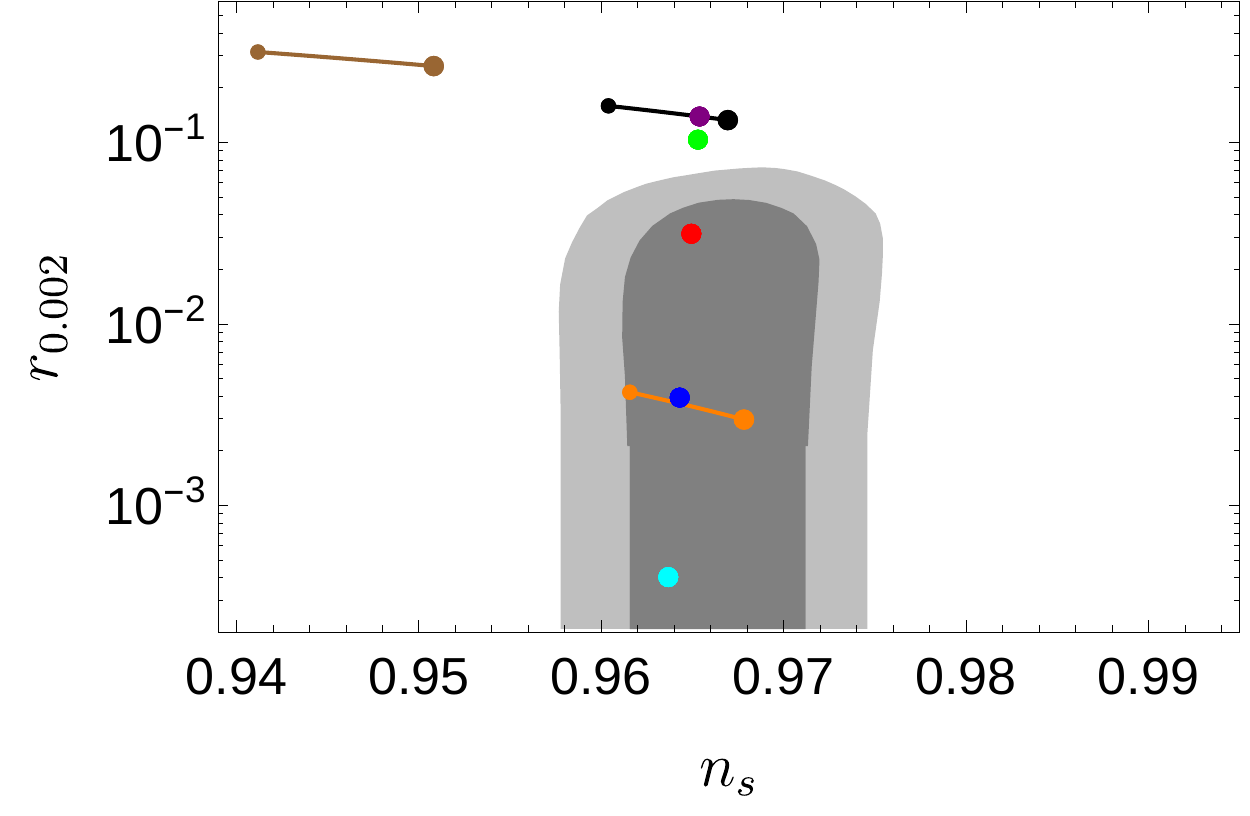}
\includegraphics[width=0.49\textwidth]{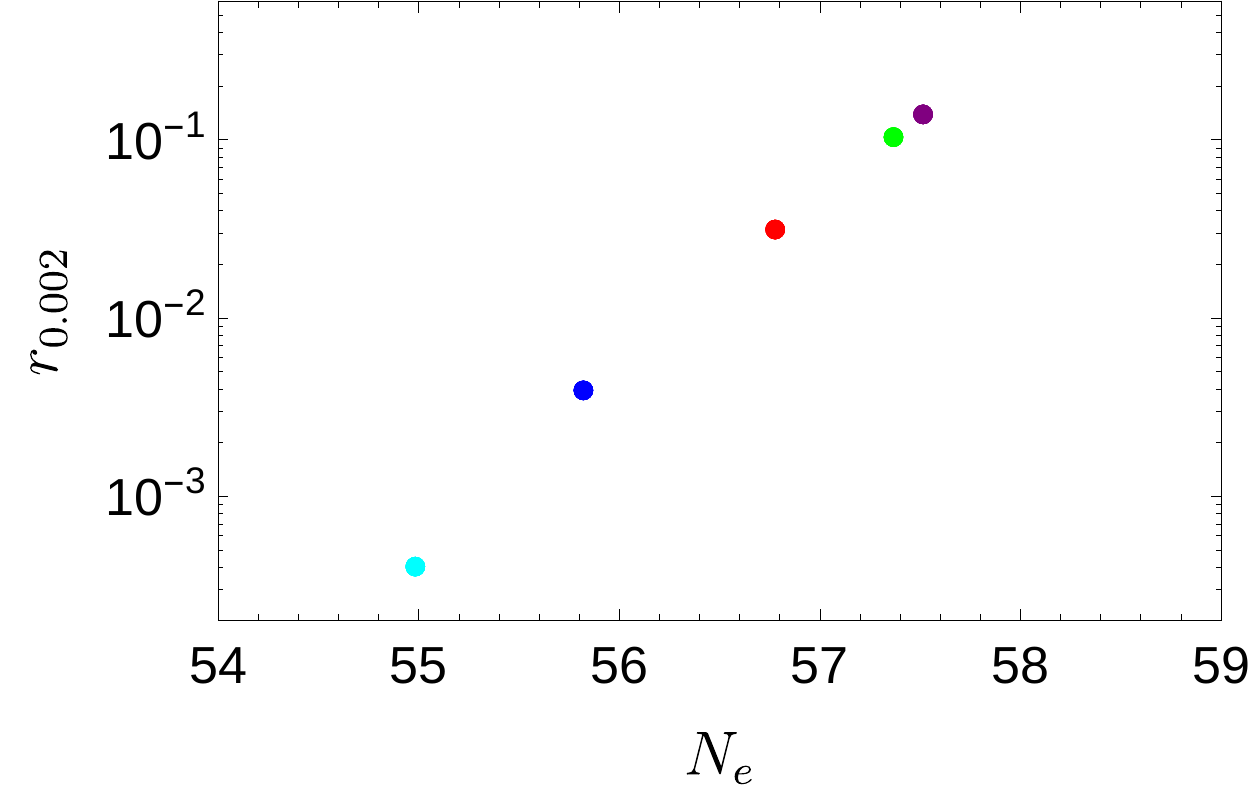}
\caption{\sf $r$ vs. $n_s$ (left) and $r$ vs. $N_e$ for the 2nd order CW  potential with $R^2$ in Palatini gravity with $\alpha=0$ (purple), $10^7$ (green), $10^8$ (red), $10^9$ (blue) and $10^{10}$ (cyan). The continuous line represents positive field ($\bar\zeta >0$ \ie $\phi > v$) inflation, while the dashed one negative field ($\bar\zeta <0$ \ie $\phi < v$) inflation. For reference, we also plot the predictions of quartic (brown), quadratic (black), linear (yellow) and Starobinsky (orange) inflation in metric gravity.  The light gray areas present the 1 and 2$\sigma$ constraints from the latest Planck data.
}
\label{fig:r:vs:ns:xi:CW:2:R2}
\end{center}
\end{figure}

The results for the 2nd order CW potential are given in figure \ref{fig:r:vs:ns:xi:CW:2:R2}. Compared to the result of \cite{Tenkanen:2019wsd}, a more precise definition for the energy density at the end of inflation following \cite{Gialamas:2019nly} and a lower upper bound on $\alpha$ have been considered. In both panels of figure \ref{fig:r:vs:ns:xi:CW:2:R2} the different colors refer to the various values of the parameter $\alpha$ discussed previously. In both panels the predictions are simple points because the 2nd order CW potential in the Einstein frame without the $R^2$ term is pure quadratic as we have already discussed \cite{Kannike2015a}. Therefore, the predictions are $v$ independent as they depend only on the mass parameter given in \eqref{eq:m:CW:2}, which is constrained from the amplitude of the scalar power spectrum (see also section \ref{reh:num}). In the left panel the purple point is in agreement with the quadratic inflation predictions as it should be. For values of $\alpha$ bigger than $\sim 10^7$ the predictions are in agreement with the observational data. This lower bound for $\alpha$ complies with the bound  derived in \cite{Gialamas:2019nly} for the quadratic inflation with an $R^2$ term. Looking at the right panel we see that the expected behavior for the number of $e$-folds is achieved again, the higher the $\alpha$ the lower $N_e$ and $n_s$.

\subsubsection{Reheating temperature}\label{reh:num}
\begin{figure}[t]
\begin{center}
\includegraphics[width=0.49\textwidth]{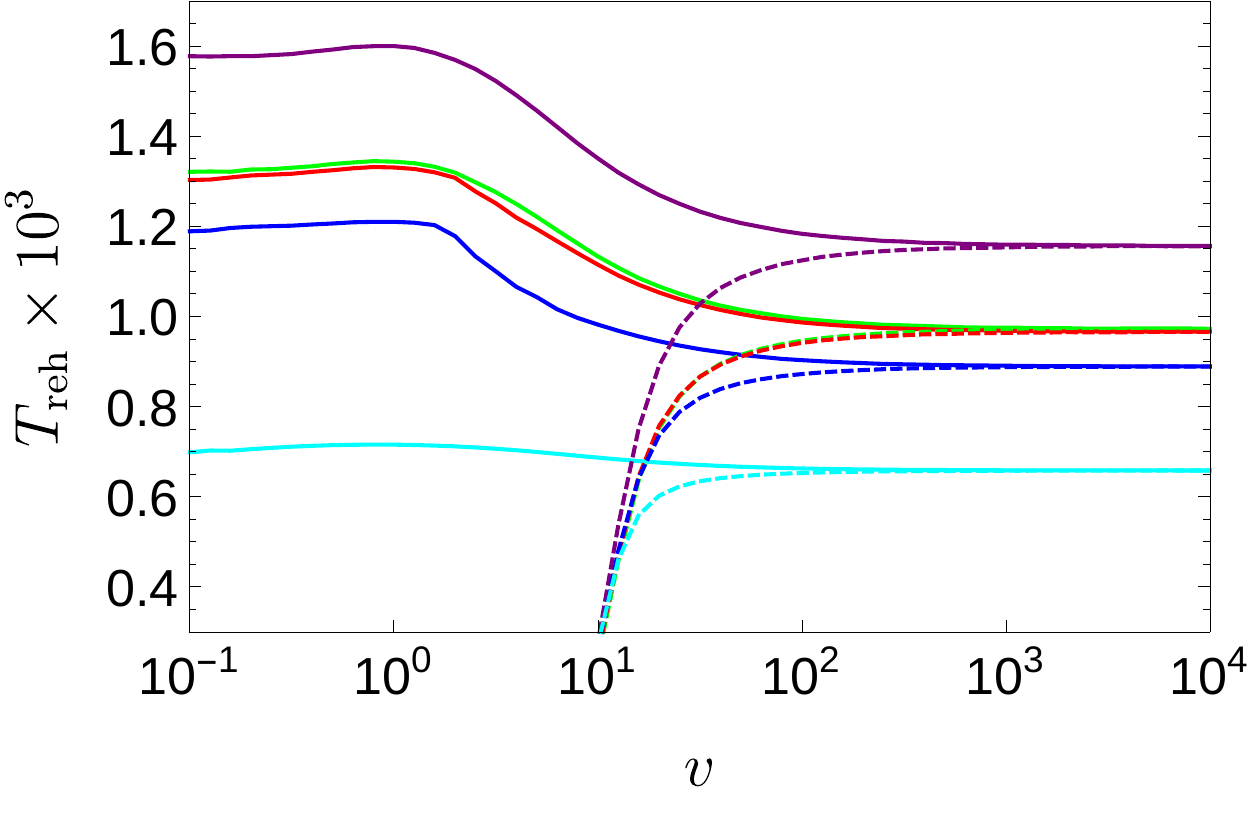}
\includegraphics[width=0.49\textwidth]{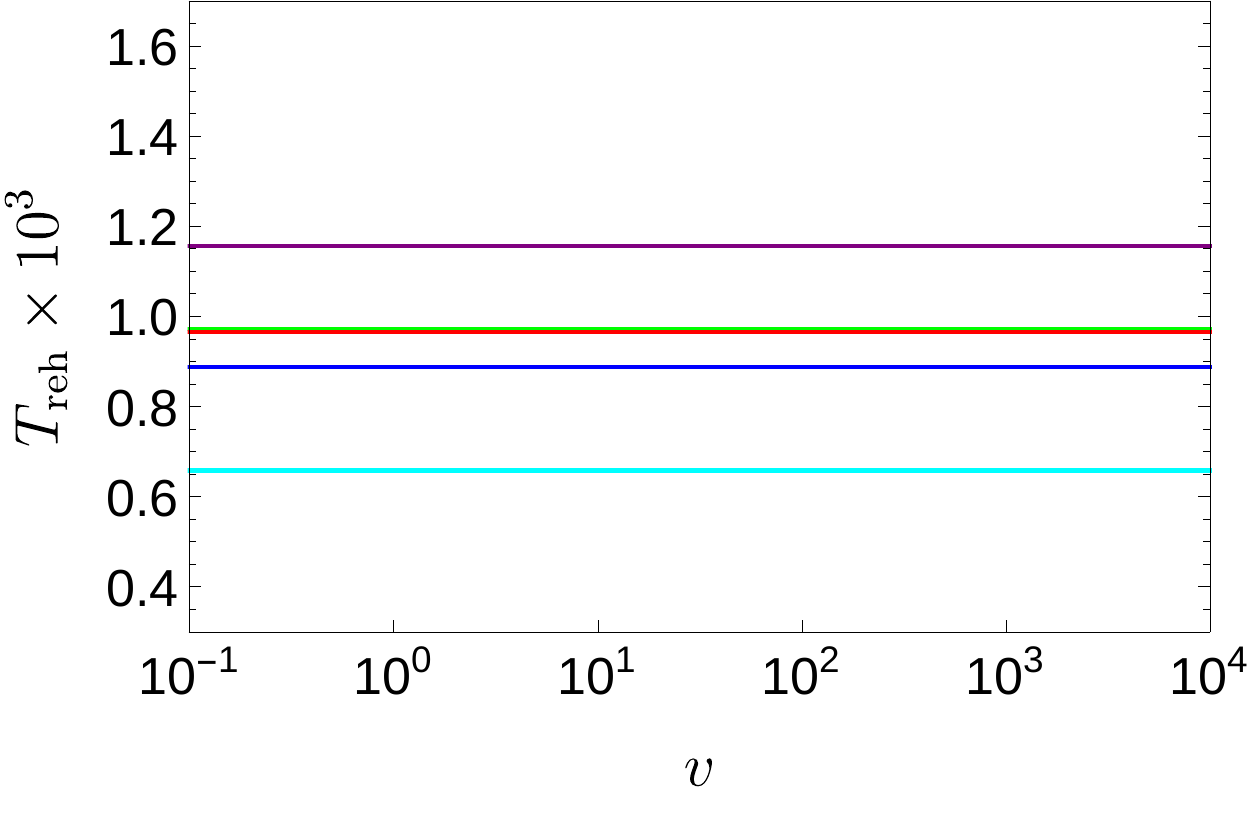}
\caption{\sf The instantaneous reheating temperature $T_{\rm reh}$ vs. $v$, both in Planck units, for the 1st order CW model (left) and for the 2nd order CW model (right) with $R^2$ in Palatini gravity with $\alpha=0$ (purple), $10^7$ (green), $10^8$ (red), $10^9$ (blue) and $10^{10}$ (cyan). The continuous line represents positive field ($\bar\zeta >0$ \ie $\phi > v$ ) inflation, while the dashed one negative field ($\bar\zeta <0$ \ie $\phi < v$) inflation.
}
\label{fig:Trehxi}
\end{center}  
\end{figure}

In figure \ref{fig:Trehxi} we display the instantaneous reheating temperature $ T_{\rm reh} $, versus the inflaton VEV $ v $, for the 1st order CW model (left) and for the 2nd order CW model (right).

For the 1st order CW model the maximum value of the instantaneous reheating temperature is achieved in the $\phi > v$ ($\bar\zeta >0$) regime. As shown, after a critical value for the $ v $ the instantaneous reheating temperature in the regime $\phi > v$ ($\bar\zeta >0$) is identified with that in the $\phi < v$ ($\bar\zeta <0$) regime, because in this region both regimes behave like quadratic inflation with an $R^2$ term.  Moreover, as we can see from the same figure the $ T_{\rm reh}$ begins to diminish (looking from left to right) at the point where $ v\simeq1 $ or equivalently at the point where linear inflation stops being a satisfactory approach of the 1st order CW model.

On the other side, for the 2nd order CW model the instantaneous reheating temperature remains constant and independent of $ v. $ This happens because in the model without the $R^2$ term the physical parameter is the mass $m_2$ given in \eqref{eq:m:CW:2}, which gets fixed\footnote{In the range $N_e \in [50,60]$ the mass parameter must be $m_2 \simeq (6.5 \pm 0.5)\times 10^{-6}$ \cite{Gialamas:2019nly} in order to comply with power spectrum data.} due to the constraint on the amplitude of scalar perturbations, $A_s \simeq 2.1 \times 10^{-9}$. Therefore the change on $ v$ is counterbalanced by a change of $  \beta'$ and the mass parameter remains the same. In the presence of the $R^2$ term the normalization of $  \beta' v^2$ remains basically unchanged (see \cite{Enckell2019}) and therefore all the physical quantities like $T_{\rm reh}$ are independent of $v$. Using the condition $\epsilon_U = 1$ to determine the field value at the end of inflation and eqs. \eqref{dens} and \eqref{Xend} to determine the energy density we can derive an analytic approximate formula for the $T_{\rm reh}$ which is too complicated to be presented, but useful to give us an estimate in the large $\alpha$ limit. In this approach $T_{\rm reh}$ is a function of the quantity $1+8\alpha   \beta' v^2$ and for $\alpha \gg 1/(8   \beta' v^2) \simeq 10^9 $ the reheating temperature has the asymptotic form $T_{\rm reh} \simeq 0.26\times \alpha^{-1/4}$ in Planck units.

Finally, comparing the plots of the two models in figure \ref{fig:Trehxi} we see that for large values of the $v$ the instantaneous reheating in the left panel converges to the one of the right panel. This happens because the 1st order CW model (without the $R^2$ term)  \cite{Kannike2016b,Racioppi2017}  behaves like quadratic inflation for large values of  $v$ (i.e small $\xi$) as we have already discussed in section \ref{subsec:xi:CW} and the 2nd order CW model (without the $R^2$ term) behaves like quadratic inflation in general as discussed in section \ref{subsec:xi:CW:2}.

\subsubsection{Validity of theory}\label{sec:validity}
We now discuss the validity of the results obtained. First of all, we remind that, contrary to its metric counterpart, the Palatini formulation of non-minimally coupled gravity does not have a unitarity problem~\cite{Bauer2011a, Shaposhnikov:2020geh, Shaposhnikov:2020fdv, McDonald2020}. Therefore the presence of a non-minimal coupling does not induce in our case a cut-off scale beyond which the theory is not anymore valid. On the other hand, we might still get some additional constraints coming from the validity of the implicit assumptions that we made when choosing the effective quartic coupling to be either \eqref{eq:lambda:run:2} or \eqref{eq:lambda:run}. 

The first assumption is that gravitational radiative corrections are negligible. Since we do not have yet a consistent and universally accepted theory of quantum gravity, we can only make a reasonable guess starting from the Einstein frame and assuming that quantum gravity effects are negligible when the inflaton energy is much below the Planck scale. Since we are dealing with slow-roll inflation, the demand is practically equivalent to requiring that the Einstein-frame inflaton potential be sub-Planckian. In particular, we need
\begin{equation}
    U_* \ll 1 \label{eq:noQG:E} \, .
\end{equation}
It is easy to check that this is always ensured because of the constraint on the amplitude of scalar perturbations $A_s \simeq 2.1 \times10^{-9}$ ~\cite{Akrami:2018odb, Ade:2018gkx}. Using \eqref{eq:As-and-ns}, we have
\begin{equation}
    U_* =  \frac{3}{2} \pi^2 A_s r \, \simeq \, 3.1 \times 10^{-8} r \, \ll 1 \, ,
\end{equation}
which ensures the consistency of the theory. Moreover, we can also check that the constraint \eqref{eq:noQG:E} protects also from quantum gravity corrections in the Jordan frame. In such a frame, with gravity being non-minimally coupled and using the slow-roll approximation, we need to require
\begin{equation}
    \frac{V_*^\text{J}}{f^2} \ll 1 \label{eq:noQG:J} \, ,
\end{equation}
where $V^\text{J}$ is the total scalar potential and $f$ is the non-minimal coupling to gravity, which can be reconstructed in our case respectively from \eqref{eq:Weylscaling} and \eqref{U2}
\begin{eqnarray}
    V^\text{J} &=& V(\phi) + \frac{\chi^2}{8 \alpha} \\
    f &=& \chi + A(\phi) \, ,
\end{eqnarray}
with $\chi$ given in \eqref{chi}. Using the slow-roll approximation we can approximate $\chi$ as 
\be
  \chi \simeq \frac{ 8 \alpha V(\phi) }{ A(\phi)} \,
\ee
insert it in \eqref{eq:noQG:J} and obtain
\begin{equation}
    \frac{V_*}{f^2} \simeq \frac{V_*}{A(\phi_*)^2+ 8 \alpha V_*} = U_* \ll 1 \, ,
\end{equation}
ensuring the validity of theory also in the Jordan frame.

\begin{figure}[t]
\begin{center}
\includegraphics[width=0.49\textwidth]{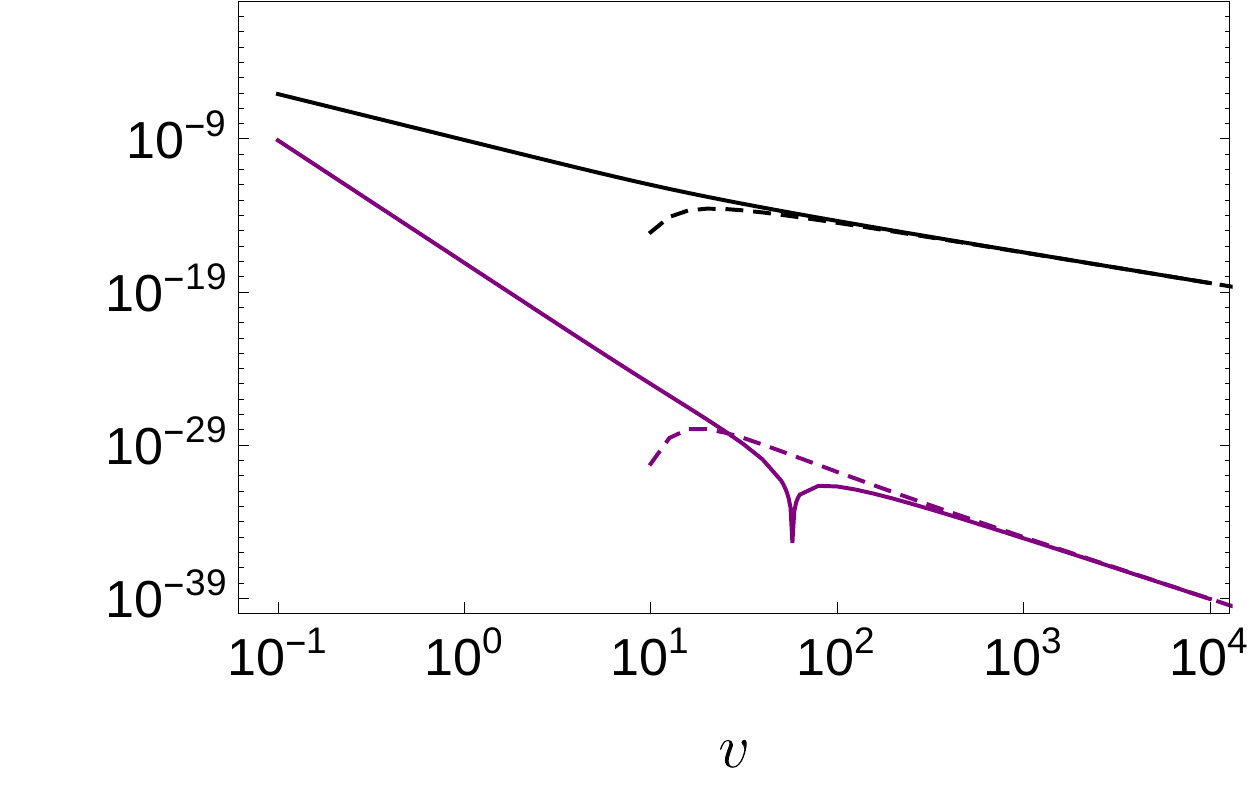}
\includegraphics[width=0.49\textwidth]{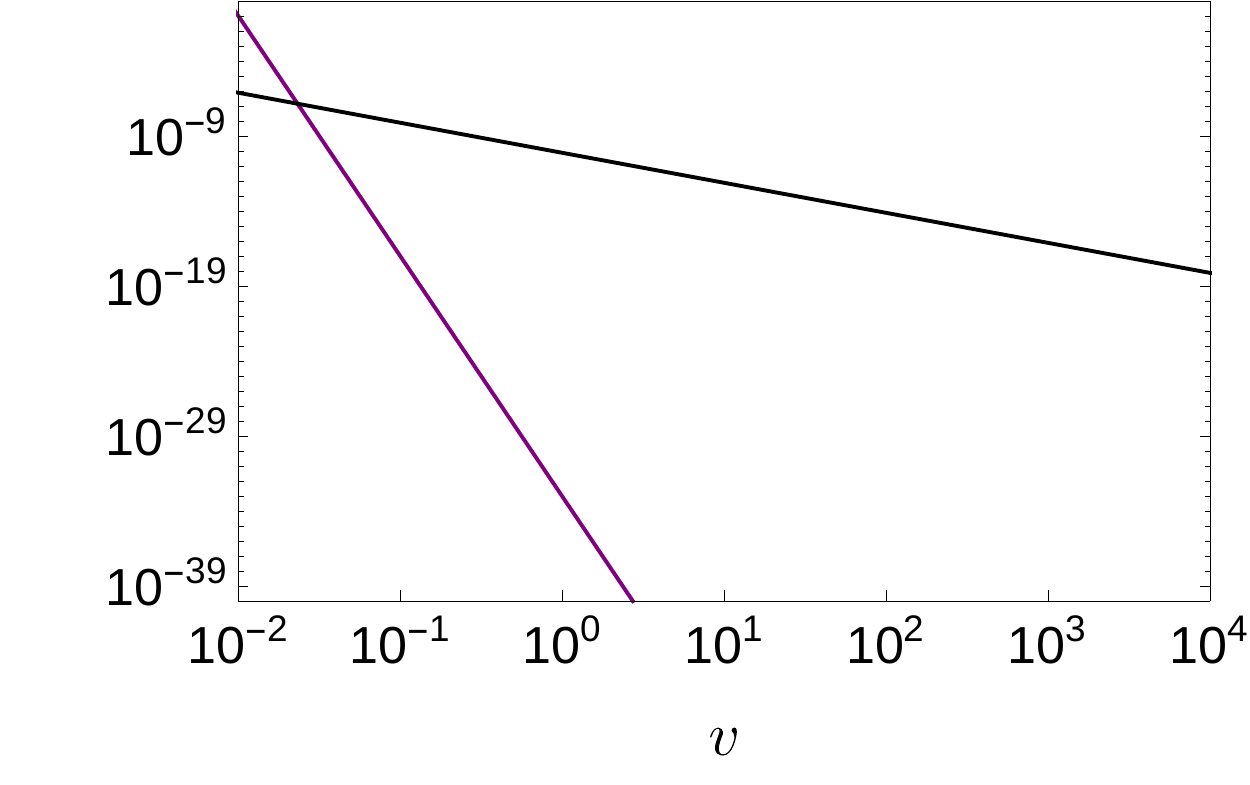}
\caption{\sf $\beta_\lambda(\phi^*)$ vs. $v$ (purple) and $\beta$ vs. $v$ (black), for the 1st order CW model (left) and $\beta_\lambda(\phi^*)^2$ vs. $v$ (purple) and $\beta'$ vs. $v$ (black), for the 2nd order CW model (right)  with $\alpha=0$. The continuous line represents positive field ($\bar\zeta >0$ \ie $\phi > v$ ) inflation, while the dashed one negative field ($\bar\zeta <0$ \ie $\phi < v$) inflation.}
\label{fig:lambdavsvev}
\end{center}  
\end{figure}

The second requirement is that the equations of the running quartic couplings [\eqref{eq:lambda:run:2} or \eqref{eq:lambda:run}] are valid at least during the duration of inflation. This is ensured by assuming the existence of an ad hoc dark sector\footnote{For some explicit realizations see \cite{Kannike2014, Kannike2015a, Kannike2016b, Kannike2017a}. Such examples are derived in the metric formulation, however the effective potential realization can be extended in a straightforward way to the Palatini one.} which provides the dominant contribution to the effective potential while the self-correction to the quartic coupling remains subdominant. The 1st order CW potential in \eqref{eq:V:CW} is linear in the logarithmic term, therefore it corresponds to a 1-loop effective potential. As such the validity of the approximation is ensured by the requirement
\begin{equation}
   \beta_\lambda(\phi) \approx \frac{\lambda(\phi)^2}{\pi^2} \ll \beta \, .
\end{equation}
On the other hand, the 2nd order CW potential in \eqref{potCW2} is quadratic in the logarithmic term, therefore it corresponds to a 2-loop effective potential. As such the validity of the approximation is ensured by the requirement
\begin{equation}
   \beta_\lambda^2 (\phi) \approx \frac{\lambda(\phi)^4}{\pi^4} \ll \beta' \, .
\end{equation}
The check of those requirements is performed in Fig. \ref{fig:lambdavsvev}, where we plot $\beta_\lambda(\phi^*)$ vs. $v$ (purple) and $\beta$ vs. $v$ (black), for the 1st order CW model (left) and $\beta_\lambda(\phi^*)^2$ vs. $v$ (purple) and $\beta'$ vs. $v$ (black), for the 2nd order CW model (right). The continuous line represents positive field ($\bar\zeta >0$ \ie $\phi > v$ ) inflation, while the dashed one represents negative field ($\bar\zeta <0$ \ie $\phi < v$) inflation. Since the normalization of the scalar potential is practically unaffected at the first order by the contribution of the $R^2$ term~\cite{Enckell2019}, we consider only $\alpha=0$, in order to make the plot more readable. For what concerns the 1st order CW model, we can see that in the considered range for $v$ the condition is always satisfied. The plot for the small field case is interrupted at $v=10$, because $v<10$ is by far out of the $2\sigma$ allowed region. We also notice that the cusp in the left panel of Fig.~\ref{fig:lambdavsvev} corresponds to the case in which $\lambda(\phi^*)=0$. On the other hand, the 2nd order CW model is valid only for $v \gtrsim 0.1$ (actually similarly to the 1st order CW case).

\section{Conclusions}\label{sec5}

The recent observational data \cite{Akrami:2018odb,Ade:2018gkx} motivated us to reconsider the previous results on the non-minimal Coleman-Weinberg inflation \cite{Kannike2016b,Racioppi2017} in presence of an $R^2$ term in Palatini gravity. In our models the Planck scale is dynamically generated from the inflaton's VEV through the non-minimal coupling between gravity and the inflaton. 

The coefficient $\alpha$ of the $R^2$ term gives us the opportunity to reduce the value of the tensor-to-scalar ratio $r\,,$ in order to be in agreement with the observational data. As $\alpha$ gets larger, without an upper limit, $r$ decreases \cite{Enckell2019}. Very large values of $\alpha$ have been used in the literature, in order to address issues like the Trans-Planckian censorship conjecture \cite{Tenkanen:2019wsd,Tenkanen:2020cvw}. However, if we want to respect the detectability of the tensor-to-scalar ratio from the next-generation CMB satelites \cite{Matsumura2016,Kogut_2011,Hanany:2019lle}, this value is restricted, $\alpha<4\times 10^{10}\,.$ 

The models under consideration have come from the general scalar potential $V(\phi) = \frac{1}{4} \lambda (\phi) \phi^4 + \Lambda^4 $, assuming that the VEV of the scalar field $\phi\,,$ is $v = \frac{1}{\sqrt\xi}\,.$ The separation of the models of section \ref{subsec:xi:CW} and section \ref{subsec:xi:CW:2} is a result of the nature of the coupling $\lambda (\phi)$ and its beta-function $\beta (\mu)= \mu \frac{\partial}{\partial \mu}\lambda (\mu)$. In the case where $\beta(v) >0$ and $\lambda(v) <0 $ \cite{Kannike2016b, Racioppi2017} the resulting potential (1st order CW potential) is asymmetrical under the field transformation ${\bar \zeta} \to -{\bar \zeta}$, where $\bar \zeta$ is the field in the Einstein frame in absence of an $R^2$ term. The asymptotic limits of the potential are interesting as they represent two common types of potentials in the Einstein frame, the linear for $v\ll 1$ and the quadratic for $v\gg 1\,.$ After the addition of the $R^2$ term the 1st order CW potential is rescued in the sense that the tensor-to-scalar ratio value is reduced. In the second case,  $\beta (v)= \lambda (v)=0$, the resulting potential \eqref{potCW2} in the Einstein frame without the $R^2$ term is purely quadratic. The physical parameter in this case is the first derivative of the beta-function times $v^2$. A quadratic potential can be also rescued with the presence of the $R^2$ term if $\alpha \gtrsim 10^7$ \cite{Antoniadis2018, Tenkanen:2019wsd, Gialamas:2019nly,  Lloyd-Stubbs:2020pvx}.

Finally, without invoking any particular reheating mechanism we have undertaken this calculation assuming instantaneous reheating, $\rho_{\rm reh}=\rho_{\rm end}$. The resulting reheating temperature for the range of parameters used and for the models studied is approximately $T_{\rm reh} \sim 10^{-3}$ in Planck units.

Eventual new measurements coming from future satellite missions \cite{Matsumura2016,Kogut_2011,Hanany:2019lle} will constrain even more the allowed region in the $r$ vs.~$n_s$ plane, reducing even more the range of allowed values for $\alpha$.

\acknowledgments

This work was supported by the Estonian Research Council grants MOBJD381, MOBTT5 and MOBTT86
and by the EU through the European Regional Development Fund
CoE program TK133 ``The Dark Side of the Universe''. The research work of I.D. Gialamas was supported by the Hellenic Foundation for Research and Innovation (H.F.R.I.) under the ``First Call for H.F.R.I. Research Projects to support Faculty members and Researchers and the procurement of high-cost research equipment grant'' (Project Number: 824).

\appendix
\section{More details about frame transformations} \label{appendix}
Our starting point is action \eqref{action1}. In order to perform comparisons between the cases $\alpha=0$ and $\alpha \neq 0$, it is convenient to look at the theory in the frame in which we would immediately recover Einstein gravity for $\alpha=0$. We call this frame ``the intermediate frame''. 
Before looking at the details of such a new frame, let us remind how a Weyl transformation acts on the curvature tensors in the Palatini formulation. In this case the connection $\Gamma^\lambda_{\rho\sigma}$ and the metric  $g_{\mu\nu}$ are treated as independent variables, therefore the Riemann tensor $R^{\lambda}_{\ \, \mu\nu\sigma}(\Gamma, \partial\Gamma)$, being constructed from $\Gamma$ and its first derivatives, is invariant under any transformation of the sole metric. The same holds for the Ricci tensor which is built by contraction of the Riemann tensor with a Kronecker delta:  $R_{\mu\nu}(\Gamma, \partial\Gamma)={\delta^\nu}_\lambda R^{\lambda}_{\ \, \mu\nu\sigma}(\Gamma, \partial\Gamma)$. On the other hand, the curvature (Ricci) scalar $R=g^{\mu\nu} R_{\mu\nu}(\Gamma, \partial\Gamma)$ is explicitly dependent on the metric and therefore under a rescaling of the metric,
\be
  g_{\mu\nu} \rightarrow \Omega^2 g_{\mu\nu} \, , \label{eq:Weyl}
\ee
$R$ scales inversely,
\be
   R \rightarrow \frac{R}{\Omega^2} \label{eq:R:scaling} \, .
\ee
Therefore it is easy to check that the action term $\int\dd^4 x \sqrt{-g} R^2$ is invariant under the Weyl transformation \eqref{eq:Weyl}. Using such a property, we can move to the intermediate frame via a Weyl scaling with $\Omega^2=A(\phi)$, obtaining
\be \label{eq:S:intermediate}
  S = \int\dd^4 x \sqrt{-g} \left[ \frac{1}{2} R + \frac{\alpha}{2} R^2 - \frac{1}{2} \left( \partial {\bar \zeta} \right)^2  - \bar{U}(\bar{\zeta}) \right]
\ee
where $\bar{U}(\phi) \equiv \frac{V(\phi)}{[A(\phi)]^2} $ is the same potential defined in \eqref{Ub} and the field redefinition is given by solving
\be \label{eq:bar:zeta:general}
  \left(\frac{d \phi}{d \bar\zeta}\right)^2 = A(\phi) \, .
\ee
Note that by imposing $\alpha=0$ we recover immediately the usual Einstein frame action. By replacing the $R^2$ term with the auxiliary field action, we obtain
\be \label{eq:SJ:intermediate:chi}
   S = \int\dd^4 x \sqrt{-g} \left[ \frac{1+\chi}{2} R  - \frac{1}{2} \left( \partial {\bar \zeta} \right)^2 - \frac{\chi^2}{8 \alpha} - \bar{U}(\bar{\zeta}) \right] \, .
\ee
Performing now an additional Weyl transformation with $\Omega^2=1+\chi$ we obtain 
\be \label{eq:SE:intermediate:chi}
   S = \int\dd^4 x \sqrt{-g} \left[ \frac{1}{2} R  - \frac{1}{2 \left(1+\chi\right)} \left( \partial {\bar \zeta} \right)^2  - \frac{1}{\left(1+\chi\right)^2} \left( \frac{\chi^2}{8 \alpha} + \bar{U}(\bar{\zeta}) \right) \right] \, .
\ee
By solving the equation of motion for $\chi$ we obtain
\be \label{eq:chi:EOM}
  \chi = \frac{ 8 \alpha \bar{U}(\bar{\zeta}) + 2 \alpha \left( \partial \bar{\zeta} \right)^2 }{ 1 - 2 \alpha \left( \partial \bar{\zeta} \right)^2 } \ ,
\ee
which is in agreement with eq.~\eqref{chi}. Inserting eq.~\eqref{eq:chi:EOM} into \eqref{eq:SE:intermediate:chi}, we get
\be \label{eq:SE}
  S = \int\dd^4 x \sqrt{-g} \left[ \frac{1}{2} R - \frac{1}{2} \left( \partial \zeta \right)^2 + \frac{\alpha}{2} \left( 1 + 8 \alpha \bar{U}(\zeta) \right) \left( \partial \zeta \right)^4 - U(\zeta)  \right] \ .
\ee
where $U \equiv \frac{\bar{U}}{1 + 8 \alpha \bar{U}}$ is the same potential as eq.~\eqref{Ueff} and $\zeta$ is defined by
\be
  \left(\frac{\dd \bar{\zeta}}{\dd \zeta}\right)^2 = 1 + 8 \alpha \bar{U} \ . \label{eq:zetabar:zeta}
\ee
The action \eqref{eq:SE} is the same as action \eqref{action6}. By using the chain rule we can also make explicit the relation between $\zeta$ and $\phi$:
\be 
  \left(\frac{\dd \phi}{\dd \zeta}\right)^2 = 
  \left(\frac{\dd \phi}{\dd \bar{\zeta}} \frac{\dd \bar{\zeta}}{\dd \zeta} \right)^2 = A(\phi) \left(1 + 8 \alpha \bar{U} \right) \,, 
\ee
in perfect agreement with \eqref{eq:phi:general}.

\bibliography{References}{}

\end{document}